%% file: ns_masses.tex
\documentclass[iop]{emulateapj}

\usepackage{amsmath}
\usepackage{amssymb}
\usepackage{amstext}
\usepackage{apjfonts}
\usepackage{graphicx}
\usepackage{epsfig}
\usepackage{color}
\usepackage{natbib}
\begin{document}

\shortauthors{Schwab et al.}
\shorttitle{Neutron Star Mass Distribution}
\title{Further Evidence for the Bimodal Distribution of Neutron Star Masses}

\author{J. Schwab\altaffilmark{1,2}, Ph. Podsiadlowski\altaffilmark{3}, and S. Rappaport\altaffilmark{1}}

\altaffiltext{1}{37-602B, M.I.T. Department of Physics and Kavli
 Institute for Astrophysics and Space Research, 70 Vassar St.,
 Cambridge, MA, 02139; sar@mit.edu,jschwab@mit.edu}
\altaffiltext{2}{Department of Physics, University of California, 
366 LeConte Hall, Berkeley, CA 94720; jwschwab@berkeley.edu}
  \altaffiltext{3}{Department of Astrophysics, University of 
Oxford, Oxford OX1 3RH, UK;  podsi@astro.ox.ac.uk}

\begin{abstract}
We use a collection of 14 well-measured neutron star masses to strengthen the case 
that a substantial fraction of these neutron stars was formed via electron-capture
supernovae (SNe) as opposed to Fe-core collapse SNe.  The e-capture SNe are
characterized by lower resultant gravitational masses and smaller natal kicks,
leading to lower orbital eccentricities when the e-capture SN has led to the formation of
the second neutron star in a binary system. Based on the measured masses and 
eccentricities, we identify four neutron stars, which have a mean post-collapse 
gravitational mass of $\sim$$1.25$ $M_\odot$, as the product of e-capture SNe.
We associate the remaining ten neutron stars, which  have a mean mass of $\sim$1.35 $M_\odot$,
with Fe-core collapse SNe. If the e-capture supernova occurs during the formation of the 
first neutron star, then this should substantially increase the formation
probability for double neutron stars, given that more systems will
remain bound with the smaller kicks. However, this does not appear to
be the case for any of the observed systems, and we discuss possible
reasons for this.

\end{abstract}

\keywords{stars: neutron, stars: evolution}

\section{Introduction}

Precise neutron star mass determinations, coupled with a theoretical knowledge of the 
pre-collapse mass, can be used to test the neutron star equation of state. 
\cite{Podsi05} used the double pulsar system J0737-3039 for such a test. They 
inferred from the low mass ($1.249 \pm 0.001 \, M_\sun$) of Pulsar B 
in this system, that it formed via an electron-capture supernova (SN), an
inference first made in \citet{Podsi04} and independently by \citet{vdH04}.
In addition to such uses, well-measured neutron star masses are extremely helpful in
understanding the formation scenarios of these objects.

In this work, we consider the entire sample of known neutron stars
with well-measured masses.  We find that the mass distribution is most
compatible with the existence of two distinct populations, a higher-mass
($\sim$$1.35 \, M_\odot)$ and a lower-mass ($\sim$$1.25 \,
M_\odot$) population.  We interpret these two populations to be the result of
distinct evolutionary formation scenarios: the low-mass population
originates in electron-capture SNe and has received low kicks, while
the high-mass population is the result of iron core collapse SNe.

In \S2 we compare and contrast the two principal channels for the
production of neutron stars: e-capture supernovae and Fe core-collapse
supernovae.  The current sample of 14 well-measured neutron
stars is presented and discussed in \S3; these all have mass uncertainties
of $\lesssim 0.025 \, M_\odot$.  In \S4 we perform some statistical tests
which provide support for the hypothesis that there are two parent
populations of pre-collapse core masses.  We carry out a simple population
synthesis study in \S5 of the expected eccentricity distributions for e-capture
and Fe core-collapse SNe, using their different anticipated core masses and natal
kick speed distributions.  In \S6, we attempt to fit the observed systems with
well determined neutron star masses into the two principal evolutionary scenarios: the
``standard'' channel and the double-core channel.  We summarize our results
and draw some general conclusions in \S7.  In particular, we find that (i)
a substantial fraction of neutron stars are formed in e-capture SNe, and (ii)
 there is evidence from our work that the double-core formation scenario
is less unlikely than previously thought by most workers in the field.

\section{Evolutionary History}

Neutron stars are believed to form through two main evolutionary
channels: iron core collapse and electron-capture supernovae. The
first occurs in a massive star when it has developed an iron core 
which exceeds the Chandrasekhar mass and 
no more nuclear burning can take place. The resulting mass of
the neutron star depends not only on the neutron star equation of
state, but also on the mass of the iron core and the maximum iron core
mass for which a successful supernova can occur. The latter depends on
the details of the supernova mechanism that are still not fully
understood. If the iron core mass is too large, the explosion
mechanism fails and the core collapses to a black hole. In the
presently most popular paradigm of delayed neutrino-driven explosions
\citep[see, e.g.,][]{Mezzacappa07, Janka08}, 
the explosion takes place when enough neutrino energy has been deposited 
in the gain region outside the proto-neutron star to
overcome the binding energy of the remaining core, stop the accretion
and initiate an outflow. The characteristic energy of such a delayed
explosion has to be of the order of the characteristic binding energy
of the remaining core ($\sim 10^{51}$\,ergs). Hence, successful
iron-core collapse supernovae are expected to have explosion energies
close to this characteristic energy.

In contrast, an electron-capture (e-capture) supernova occurs in a
very degenerate ONeMg core, long before an iron core has developed,
and is triggered by the sudden capture of electrons onto Ne
nuclei, taking away the hydrostatic support provided by the degenerate
electrons \citep[e.g.,][]{Nomoto84}. This occurs at a characteristic density
($\sim 4.5\times 10^9$\,g\,cm$^{-3}$; \citealt{Podsi05}), which
in turn can be related to a critical pre-collapse mass for the ONeMg core of 
$\sim 1.37\,M_\odot$. Hence, an e-capture supernova is expected to occur
when a degenerate ONeMg core reaches this critical mass either by
accretion from an envelope (inside an AGB star [e.g., 
\citealt{Siess07, Poelarends08}] 
or in a helium star [e.g., \citealt{Nomoto87}]), by accretion
from a companion star (so-called accretion-induced collapse [e.g.,
\citealt{Nomoto91}]), or as a consequence of the merger of two CO
white dwarfs and the subsequent formation of an ONeMg core 
\citep[e.g.,][]{Nomoto85}.  Since the collapse occurs at a characteristic
ONeMg core mass, the resulting neutron-star mass is entirely
determined by the equation of state and the amount of core material
that is ejected in the supernova.\footnote{This ignores the role of
 rotation which may be important, in particular, in the case of an
 accretion- or merger-induced collapse.} The case of Pulsar B in the
double pulsar system J0703-3039 suggests that this mass is close to
$1.25\,M_\odot$ \citep{Podsi05}. Furthermore, since essentially the
whole core collapses to form a neutron star, the remaining envelope is
relatively easy to eject, leading to a fainter supernova with the
ejection of very few heavy elements \citep[see, e.g.,][]{Dessart06,Kitaura06}.
It has recently been argued that the large kicks
most neutron stars receive at birth (Hobbs et al.\ 2005) are caused by
an accretion shock instability that causes a wobbling of the core,
imparting momentum in the process 
(e.g.,\,\citealt{Blondin06, Blondin07, Foglizzo07}; but see \citet{Fryer07} for
a more skeptical point of view).  Since, in the case of an e-capture supernova, 
the explosion occurs before these instabilities have time to grow, no large kick 
is expected for a neutron star formed through this channel.

The suggestion that electron-capture supernovae may
produce low supernova kicks and a distinct low-mass neutron-star
population was first made independently by \citet{Podsi04} and
\citet{vdH04}.\footnote{The latter author also suggested a
third more massive population of neutron stars with masses around
1.85\,$M_\odot$ from stars with an initial mass around 20\,$M_\odot$.}
 \cite{vdH04} specifically discussed this low-mass, low-kick
 population in the context of binary radio pulsars, 
and used the then-current observations of several NS-NS
binaries and a NS-WD binary to argue that they formed
via e-capture.

\begin{deluxetable}{lll}[h]

\tablecolumns{3}

\tablecaption{Comparison of Fe-Core Collapse and e-Capture Supernovae}

\tablehead{
	\colhead{Properties} &
	\colhead{Iron Core Collapse} &
	\colhead{e-Capture SN}  \\
	} 
\startdata
{\em Supernova Properties} & & \\
Explosion energy & $\sim 10^{51}\,$ergs & $\la 10^{50}\,$ergs \tablenotemark{a} \\
Ejecta & rich in heavy elements & few heavy elements \\
& (Fe, Si, O) & \\
&& \\
{\em Neutron Star Properties} & & \\
Masses & range of masses & characteristic mass \\
&& $\simeq 1.25\,M_\odot$ \\
Neutron star kick & large standard kick & low kick \\
& ($\sigma\simeq 265\,$km\,s$^{-1}$) \tablenotemark{b} & \\
&& \\
{\em Binary Properties} & & \\
Occurrence & single or binaries & preferentially \\
& & in binaries \tablenotemark{c} \\
Eccentricity & high & low \\
Recycled pulsar spin & mis-aligned with orbit & aligned with orbit  \\
& (e.g.\ geodetic precession) & \\
\enddata 
\tablenotetext{a}{\citet{Dessart06,Kitaura06}}
\tablenotetext{b}{\citet{Hobbs05}}
\tablenotetext{c}{\citet{Podsi04}}
\end{deluxetable} 

Table~1 summarizes the main differences in neutron-star and supernova 
properties for these two channels. Note, in particular, that, for
neutron stars formed from iron-core collapse, one expects a range
of masses that is determined by the range of iron core masses in the
progenitors that allows a successful explosion, while in the case
of neutron stars from an e-capture supernova one expects a fairly
well determined mass. Thus, the distribution of post-supernova
neutron star masses directly constrains not only the equation of state, but
also the properties of successful supernova explosions.

\section{Neutron Star Sample}

\begin{figure}[h]
\vspace{0.3cm}
\begin{center}
\includegraphics[width=0.48 \textwidth]{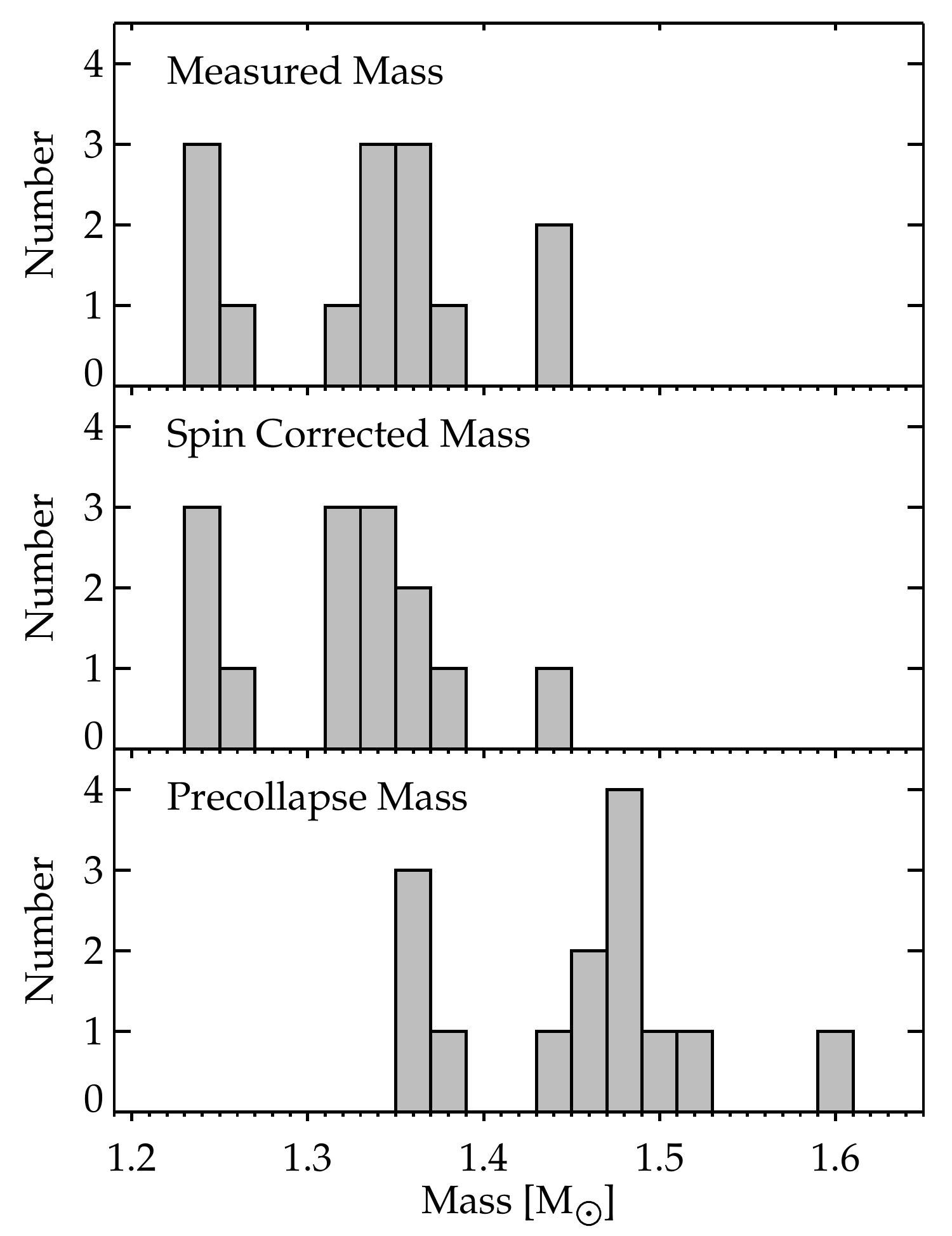}
\caption{Mass histograms for the sample of 14 neutron stars. 
({\em Top Panel}): The measured (gravitational) masses of the neutron stars.
({\em Middle Panel}): The masses of the neutron stars corrected for accretion as discussed in the text.
({\em Bottom Panel}): The precollapse (baryonic) masses of the neutron stars, based on one particular illustrative 
neutron-star equation of state.}
\label{fig:hist}
\end{center}
\end{figure}

There are 14 neutron stars which have masses known with an accuracy of better than $\sim$
0.025 $M_\odot$. The majority of these  (twelve) are from double neutron star systems; two are in 
binary systems with suspected white dwarf companions. The properties of these systems are summarized in 
Table \ref{tbl:nsmass} (for references see, e.g., \citealt{Stairs08}).  A histogram of the measured 
gravitational masses is shown in the top panel of \mbox{Figure \ref{fig:hist}.}

\begin{deluxetable*}{lcccccc}[hb]

\tablecolumns{7}
\tablewidth{0pt}

\tablecaption{14 Well Measured Neutron Star Masses}

\tablehead{
	\colhead{Pulsar Name} &
	\colhead{Mass of Recycled} &
	\colhead{Mass of Young}  & 
	\colhead{$P_{\rm orb}$} &
	\colhead{Eccentricity} &
	\colhead{Pulse Period} &
	\colhead{Reference} \\
	& 
	\colhead{Neutron Star $(M_\odot)$} &
	\colhead{Neutron Star $(M_\odot)$} & 
	\colhead{(hours)} &
	& 
	\colhead{(ms)} &
	} 
	
\startdata
\input{ns_masses.tbl}
\enddata
\tablecomments{All known neutron stars with a mass measured with better than $0.025 ~M_\odot$ accuracy.}
\tablenotetext{$\dagger$}{These periods are said to be associated with the ``young pulsar''.}
\label{tbl:nsmass}
\end{deluxetable*}

The rapidly rotating pulsars have likely been spun up by the accretion of a small to modest amount 
of matter ($0.001 - 0.07\,M_\odot$). We correct for this effect by subtracting the mass which would 
be necessary to spin up the star, treating it as a classical uniform-density sphere accreting from a 
disk that extends down to its surface. We have verified that for a range of plausible equations of state for neutron 
star matter, more sophisticated treatments lead to accreted (gravitational) masses that differ from our simple 
model by less than $\sim$10\% (see, e.g., \citealt{Cook94}). The results are shown in the middle panel of 
Figure \ref{fig:hist}.  Note the high degree of 
similarity of this histogram with that for the uncorrected masses; the maximum mass correction for 
any one neutron star is $\sim$0.07 $M_\odot$ (for J1909-3744).  The corrections for the other 
neutron stars were less than $\sim$0.02 $M_\odot$.  

Finally, we used a representative equation of state for neutron star matter 
\citep*[``MPA'',][]{Muther87} to translate the observed gravitational mass into a pre-collapse mass 
by calculating the baryonic mass corresponding to each gravitational mass.  The results are shown 
in the bottom panel of Figure \ref{fig:hist}.  In general, the pre-collapse masses are shifted upward by $
\sim$0.13 $M_\odot$. 

As the equation of state remains theoretically uncertain, we calculated the 
corrections for each of the equations of state collected in \citet{Lattimer01}. Within this collection, the correction
to the mass of a 1.25 $M_\odot$ neutron star varied over the range $0.09 - 0.18$ $M_\odot$. However, given the 
small range in mass considered (1.25 - 1.4 $M_\odot$), the choice of equation of state has little effect
on the relative correction between  any two systems within this range.
The net result of choosing a different equation of state would be a systematic shift in the bottom panel of 
Figure \ref{fig:hist}, as opposed to any significant stretching or skewing.

One can see from Figure \ref{fig:hist} that there are two apparent populations of neutron star mass: one 
centered at $\sim$$1.25 \, M_\odot$ and $\sim$$1.35 \, M_\odot$ (post collapse) and $\sim$$1.37 \, M_\odot$ 
and $\sim$$1.48 \, M_\odot$ (pre-collapse -- for the assumed ``MPA'' neutron-star equation of state).  The higher of these 
two mass groups is suggestive of an origin in an Fe core collapse SN, while the lower of the two groups likely 
comes from electron-capture SN events.

\section{Statistical Tests}

We make use of two statistical tests to try to quantitatively evaluate our hypotheses, the 
Kologorov-Smirnov (KS) test and the Anderson-Darling (AD) test \citep[e.g.,][]{NR07}. 
The AD test is more powerful as it takes into 
account the integrated difference between the cumulative distributions one is comparing, while the KS test 
considers only the maximum difference.

\begin{figure}
\vglue1cm
\includegraphics[width=0.48 \textwidth]{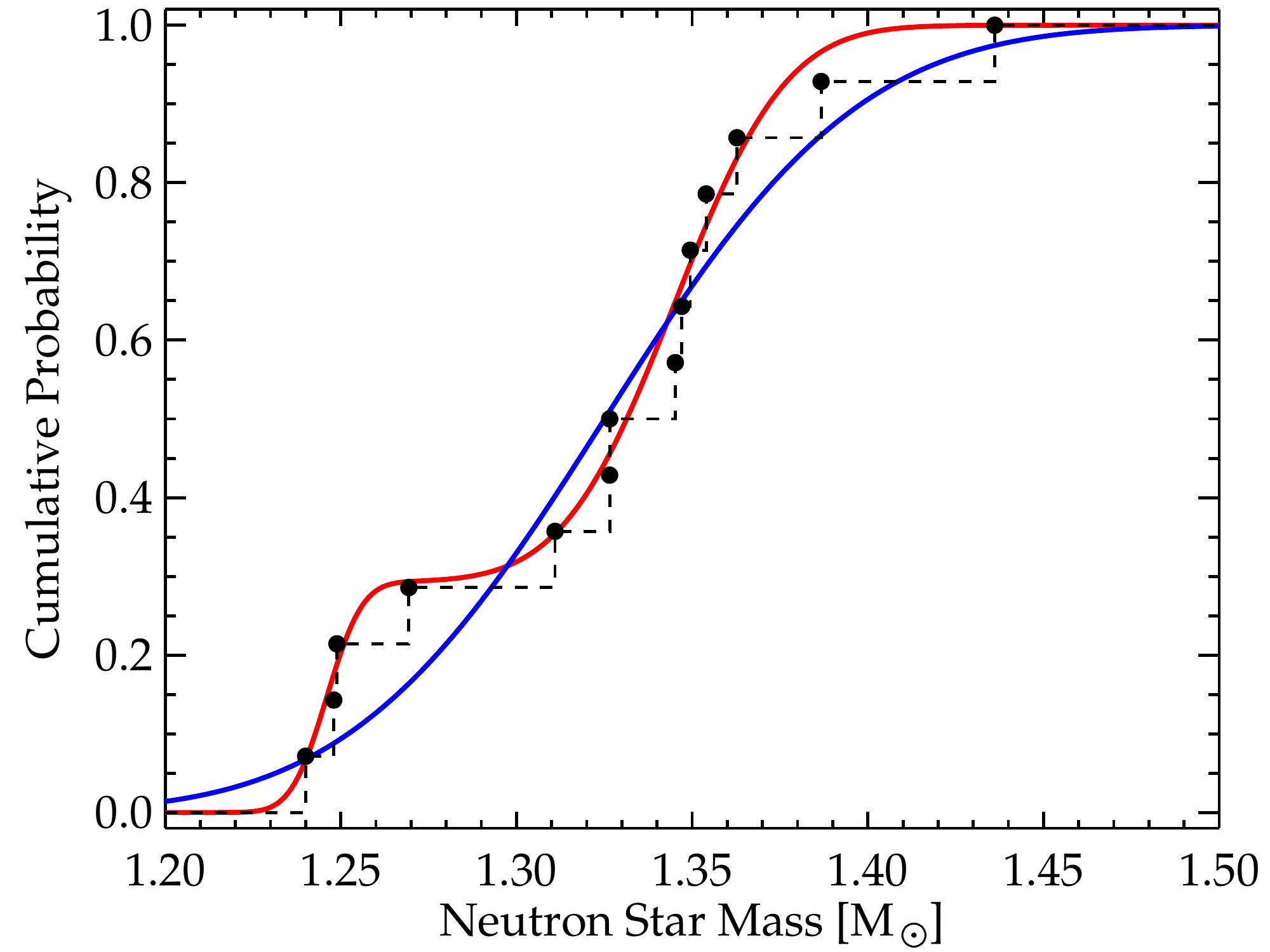}
\caption{The cumulative distribution of neutron star masses. The observed distribution is shown by 
the black dots and dashed black line. The CDF for the best two population model is shown by the 
solid red line (see text).  The blue curve is the CDF for a single Gaussian with the population mean 
and standard deviation.}
\label{fig:nsdist}
\vglue0.5cm
\end{figure}

The first test we perform is for normality, checking whether the distribution is consistent with a single
Gaussian which has the mean and standard deviation of the observed populations. The cumulative 
distribution for the observed neutron stars is shown in Figure \ref{fig:nsdist} as filled circles 
connected by a dashed histogram. The cumulative distribution for the single Gaussian described by 
a mean mass of  $1.325\,M_\odot$ and standard deviation of $0.056\,M_\odot$ is shown 
as the blue curve in Figure \ref{fig:nsdist}. We were not able to reject this hypothesis of a single 
Gaussian mass distribution with a KS test, but were able to marginally reject it at the 70\% confidence level with 
the AD test.

We also tested the hypothesis that there are two distributions present, each of which is represented by a 
Gaussian.  The best such fit is given by:
\begin{equation}
dN/dM = 0.707 \, e^{(M-1.345)^2/2\sigma_1^2} + 0.293 \, e^{(M-1.246)^2/2\sigma_2^2}~~~,
\end{equation}
with $\sigma_{1} = 0.025 \pm 0.004\, M_\odot$ and $\sigma_{2} = 0.008 \pm 0.005 \, M_\odot$.  The 
uncertainties on the mean masses of the two Gaussians are $1.345 \pm 0.003 \, M_\odot$ and
$1.246 \pm 0.003 \, M_\odot$, respectively.  The cumulative 
distribution function  corresponding to this distribution is shown plotted as a solid red curve in Figure 
\ref{fig:nsdist}.  The amplitudes of the two Gaussians are, as expected, reflective of the fact that 4 of 
the 14 neutron stars are in the lower-mass group.  As can be seen from the cumulative distribution 
for the double Gaussian, the fit is very good, compared to a single Gaussian with the population 
mean and standard deviation.  

While these statistical tests and fits do not, by themselves, constitute a proof of two populations, 
coupled with the other pieces of evidence (i.e., appropriate system eccentricities and theoretically 
expected masses for e-capture SNe), they do lend support for the hypothesis of two populations.

\begin{deluxetable*}{lllll}

\tablecolumns{5}
\tablewidth{0pt}

\tablecaption{Order of Fe-Core Collapse vs. e-Capture Supernovae}

\tablehead{
	\colhead{Category} &
	\colhead{Neutron Star Formation Type and Order} &
	\colhead{Standard Scenario} &
	\colhead{Double Core Scenario}  & \colhead{Observed?} \\
	} 
\startdata
I & Fe core collapse + Fe core collapse & possible & probable & yes  \\
& & & &\\
II & e-capture + Fe core collapse & most favored & inconsistent & no \\
& & & &\\
III & Fe core collapse + e-capture  & possible & probable  & yes\\
& & & &\\
IV & e-capture + e-capture & possible & some fine tuning & no \\
\enddata
\label{tbl:order}
\end{deluxetable*}

\section{Eccentricity Calculations}


In order to illustrate what the eccentricity distributions of e-capture vs. Fe core collapse SNe might look like, we 
carried out the following simple statistical study.  In all cases, we assume that the {\em second} SN explosion takes 
place with a He or CO core in a circular orbit with the first-born neutron star.  We take the core mass to be in the 
range $1.5 - 2.0 \, M_\odot$ for the e-capture scenario and $2.5 - 6.0 \, M_\odot$ for the Fe core collapse scenario 
\citep[e.g.,][]{Podsi04, Dewi05}. The orbit is assumed to have been circularized during a prior episode when the 
evolving core expands sufficiently to transfer at least a small amount of mass to the first-born neutron star, thereby 
spinning it up to millisecond rotation periods.  The orbital separation at the time of the second SN explosion is taken 
to be uniformly distributed over the range $1-3$ times the orbital separation needed for the He or CO core to fill its 
Roche lobe (a result expected from more detailed population synthesis calculations, e.g., \citealt{Dewi06}).  Finally, the 
natal kick distribution is taken to be a Maxwellian with $\sigma$ equal to either 30 km s$^{-1}$ or 265 km s$^{-1}$, 
for the e-capture and Fe core collapse scenarios, respectively \citep[e.g.,][]{Dewi06}.  The 
resulting orbital eccentricities are computed from the expression given by \citet{Brandt95}. The above 
assumptions should hold for both the standard formation channel as well as for the double-core channel.

\begin{figure}
\centering
\includegraphics[width=0.48 \textwidth]{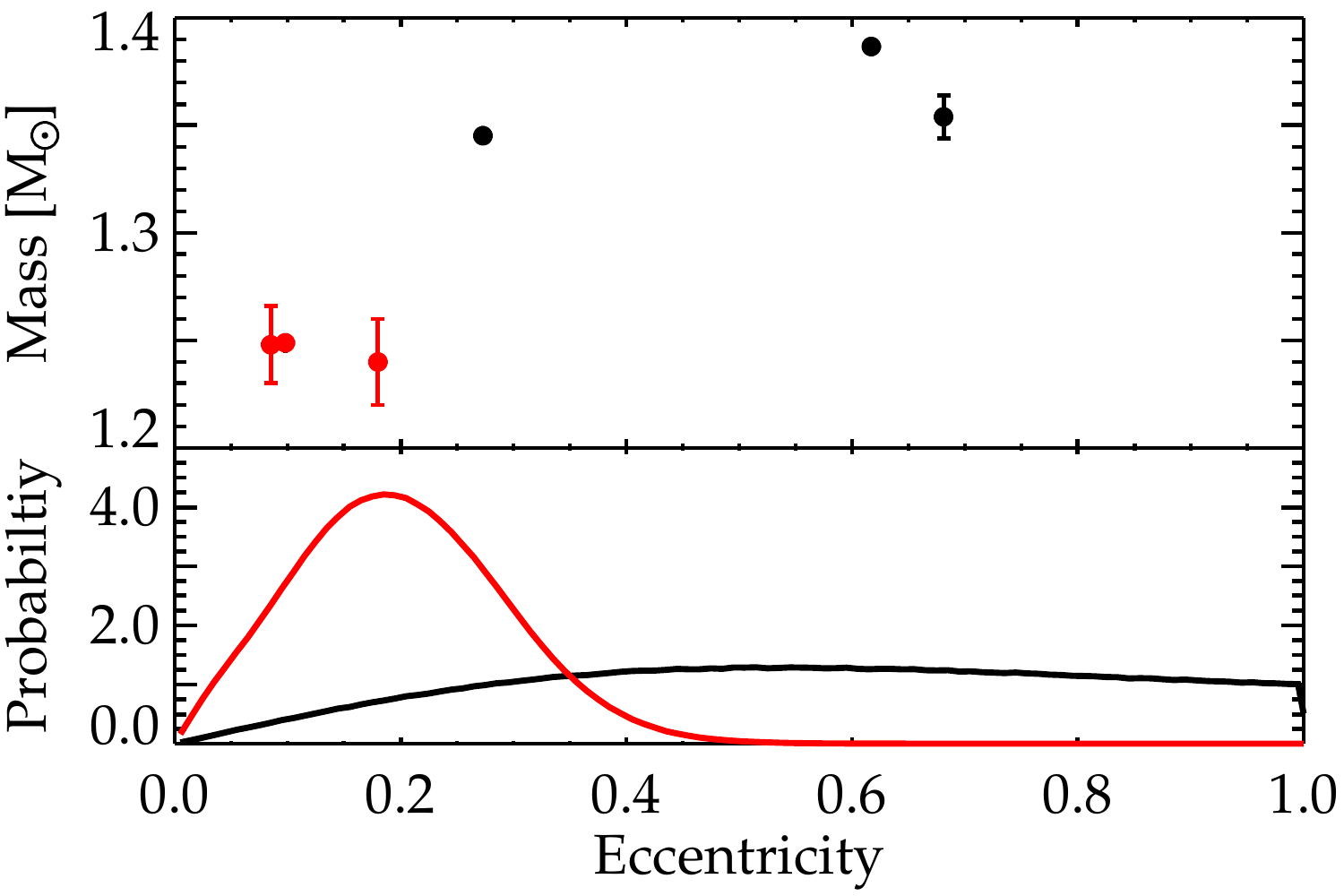}
\caption{Comparison of system eccentricities of the double neutron star binaries.
({\em Top Panel}): The measured masses of the young neutron star plotted against the system eccentricity. 
We identify the low-mass, low-eccentricity systems as being the result of e-capture SNe, marking them as red. 
An amount 0.01 was artificially added to the eccentricity of J0737 to separate it from J1906 on the plot. 
({\em Bottom Panel}): The results of the Monte Carlo eccentricity simulation described in the text. The black curve is
the distribution of Fe-core collapse systems and the red curve is for electron-capture systems.}
\label{fig:ecc}
\end{figure}

For each scenario, the initial system parameters (i.e., core mass and orbital separation) and the natal kick were 
chosen via Monte Carlo techniques for $10^7$ systems.  The results are shown in Figure \ref{fig:ecc}.  The red curve 
represents the eccentricity distribution for the e-capture scenario while the black curve is for the Fe core collapse 
explosions.  Note that in the latter case the eccentricity distribution is rather broad (in fact some $\sim$60\% of the 
systems become unbound) while for the e-capture events the eccentricity distribution peaks at $e \simeq 
0.2$, and extends down to rather low values of $e$.  In fact, we observe that these two distinct distributions 
are rather consistent, respectively, with the three cases we tentatively identify with e-capture SNe and those three 
with higher eccentricities that we associate with Fe core collapse SNe.

\section{Consistency of the Scenarios}

There are two main scenarios for the formation of double neutron star
systems.  They both start with a pair of massive primordial stars,
i.e., with masses between $\sim$8 and 25 $M_\odot$.  In the ``standard
scenario'' the more massive star evolves first, fills its Roche lobe,
and stable, quasi-conservative mass transfer to the secondary may
occur if the mass ratio is not too extreme and if the initial orbital
period is in the right range (i.e., $\sim$ a month to a year).  In
this scenario, the system can not undergo a common envelope phase
during the first stage of mass transfer, otherwise the orbit would not
be sufficiently wide after the formation of the first neutron star to
allow for the successful production of the second neutron star (as
discussed below).  Given the fact that the orbit should initially be
wide, it is actually advantageous for the first supernova explosion to
occur via an e-capture SN, in order to yield a small natal kick and
thereby help the system remain bound.  There is indeed a class of such
wide binaries containing a neutron star in nearly circular orbit with
a massive donor \citep[see, e.g.,][]{Pfahl02, Podsi04}.
After the formation of the first neutron star, the
companion star evolves, fills its Roche lobe, and the subsequent mass
transfer onto the neutron star leads to a common envelope phase due to
the extreme mass ratio of the system.  In order for the common
envelope phase to avoid a merger of the neutron star and the core of
its companion, the orbit must be wide to begin with, as alluded to
above.  The result of a `successful' common envelope phase is a He or
CO core in close ($\lesssim$ 1 day) orbit with the first neutron star.
The original neutron star is spun up via accretion from its He/CO star
companion as it evolves toward core collapse.  If the subsequent
supernova explosion is via e-capture (for He-star companions of 
mass $\lesssim 2\,M_\odot$; \citealt{Nomoto84}), then
the natal kick may be small and the final binary pair of neutron stars
would have a modestly small eccentricity and low systemic space
velocity \citep{vdH07}.  Whether the second neutron star would form via an e-capture
SN or Fe core-collapse SN would depend on the original mass of the
secondary star and the orbital period after the first neutron star has
been formed.

Both this standard scenario and the ``double core'' scenario that we describe next 
are summarized in schematic diagrams in \citet[][Fig.\,1]{Podsi05} and 
\citet[][Fig.\,1]{Dewi06} The various possible orderings for the different core-collapse 
mechanisms for the two neutron stars in the standard scenario are summarized in 
Table \ref{tbl:order}, along with some qualitative comments on the likelihood of each.
In particular, in a recent detailed population synthesis study of DNSs, 
which included both the standard and the double-core
channel, \citet{Belczynski09} predicted that category II 
systems should be by far the most dominant systems.
From the above discussion and the notes in Table \ref{tbl:order}, we would expect 
double neutron stars formed via the  standard scenario to consist of a recycled pulsar 
with a mass indicative of an e-capture SN formation (Categories II and IV; i.e., $\sim$
$1.25\,M_\odot$).  Moreover, if the orbital eccentricities are low, then both neutron 
stars would most likely be formed via e-capture SNe (i.e., Category IV).  Neither of 
these expectations is borne out by the observational data in Table 2.  

In the ``double-core'' scenario \citep{Brown95, Bethe98, Dewi06}, 
the primordial binary is required to have a pair of
stars whose mass is the same to within $\sim$$3-7\%$ of each
other. Given that massive stars seem to often reside in binaries whose
stars have comparable mass, this $\sim$5\% ``window'' is not nearly as
rare as it might seem.  In fact, roughly speaking some several percent 
of all massive stars may occur in such comparable-mass binaries if the binary 
fraction is high and if the mass ratio distribution is roughly flat.  If the two
stars have comparable mass, and the orbit is relatively wide (i.e.,
$\sim$months to years) then both stars will enter a double
common-envelope phase once the primary starts to transfer mass to the
secondary.  At that point the primary is expected to have evolved a CO
core while the secondary will also be evolved, but will more likely have 
only a He core.  Both of these cores will spiral in inside the common envelope
formed of the envelopes of both stars.  The result will be a close pair
of a CO and a He core (or less likely a pair of CO cores).  The
orbital period of the primordial binary should be wide (i.e., months
to years) if the cores are to avoid merger during the common-envelope
phase.  The CO core evolves first, most likely, to an Fe core-collapse supernova 
and leaves a $\sim$1.35 $M_\odot$ neutron star (i.e., Categories I and III; see
Table \ref{tbl:order}), though it could also experience an electron-capture SN
(Category IV).  When the He core evolves,
it expands somewhat and transfers some mass to the first neutron star,
thereby recycling it.  If the He core mass is relatively low
($\lesssim 2 \,M_\odot$; \citealt{Nomoto84}),  it will go on to undergo an e-capture
supernova, leaving a lower-mass neutron star with a smaller natal kick (i.e., Category III).
Thus, in this scenario, the
recycled pulsar is the more massive of the two, having been formed in
an Fe core-collapse supernova, while the second, an unrecycled pulsar,
should be lower in mass.  

This double-core scenario is consistent with the data for the six
neutron star binaries.  Three of them have higher orbital
eccentricities, $e = 0.27, 0.62, 0.68$, with both neutron stars in the
system having masses consistent with Fe core-collapse supernovae (Category I).
The other three have lower eccentricities, $e = 0.08, 0.09, 0.18)$, and
the unrecycled pulsar is consistent with having undergone an e-capture
supernova (with mass $\sim$$1.25\,M_\odot$), i.e., Category III (see also \citealt{vdH07}).
The eccentricities and NS masses of these six systems are plotted in Figure \ref{fig:ecc}.
Since we have only six systems, it is premature to
draw any firm conclusions, though it is remarkable that
none of the observed systems falls into the a priori most
favored standard scenario category.

In the case of the two neutron star-white dwarf binaries, the scenario is likely 
somewhat different than described above.  One of the neutron star--white dwarf 
systems (J1141-6545) has a significant eccentricity of 0.17 and a lower-mass neutron 
star. This suggests that, in this case, the white dwarf actually formed before the 
neutron star, otherwise the second phase of mass transfer, leading to the formation of 
the white dwarf, would have circularized the orbit.  This is the likely consequence of a 
first phase of conservative mass transfer where the initial masses of the binary 
components were relatively close and the primary had a mass just below the 
minimum main-sequence mass for neutron-star formation ($\sim 7\,M_\odot$). After 
this first mass-transfer phase, the secondary has accreted enough mass to end its 
evolution as a neutron star \citep[see, e.g.,][]{Church06}. In addition, after the 
ensuing common envelope phase, the 
secondary is more likely to have a mass just above the minimum neutron star 
formation mass and this naturally favours an e-capture collapse.  
The formation of this system via e-capture was previously suggested by \citet{vdH04,vdH06}.
The other 
neutron star--white dwarf system (J1909-3744) is highly circularized and it is likely 
that the neutron star formed first and the orbit was circularized during a common 
envelope phase involving the progenitor of the white dwarf.

\section{Summary and Conclusions}

We have shown that the population of 14 well-measured neutron-star masses is consistent with being comprised 
of a subset of four that were likely the result of electron-capture SNe, while the others resulted from Fe-core 
collapse SNe.  The lower neutron star masses ($\sim$$1.25\,M_\odot$) of the four candidate electron-capture SNe
and relatively low orbital eccentricities ($e \lesssim 0.18$) of the systems that contain them are both good indicators 
of this type of formation mechanism.  The remaining 10 neutron stars have larger masses ($\sim$$1.35\,M_\odot$) 
more indicative of Fe core-collapse SNe; in the systems where these more massive neutron stars formed
second, larger orbital eccentricies are observed, consistent with a larger natal kick.

We discussed four categories of formation scenarios for producing neutron stars in close binaries, especially double 
neutron stars.  These include the possibilities that (i) either the first or second neutron star was formed via an Fe core 
collapse SN and (ii) that either neutron star might have been formed in an electron-capture SN.  Any of these four 
possibilities could be connected with either the ``standard scenario'', in which a common envelope phase occurs only 
{\em after} the first SN explosion, or with the so-called ``double-core'' scenario, wherein both comparably massed 
primordial stars are simultaneously stripped of their envelopes.  (See Table~\ref{tbl:order} for a summary of the eight 
possible combinations.)

None of the observed systems falls into Categories II or IV, because the recycled pulsar is never the
lower-mass product of an e-capture supernovae. For producing double neutron star systems,
it appears that the standard scenario is somewhat disfavored. The evidence that we have presented
and examined seems to favor Category I (both the young and the recycled NS are higher-mass)
and Category III (a recycled higher-mass NS and a young lower-mass NS), both forming via the 
double-core channel (see Table~\ref{tbl:order}).

If our interpretation of the neutron star mass data is correct, i.e.,
that the double-core channel is the preferred scenario for producing
double neutron stars, this has profound implications for the fate of
neutron stars within a common envelope.  Recall that the original
motivation for proposing the double-core channel was the estimation,
by \citet{Chevalier93, Chevalier96} that a neutron star within a
common envelope might undergo hypercritical accretion and collapse to
a black hole. On the other hand, there is evidence that at least
some neutron stars survived a common envelope without being
converted into a black hole (e.g., PSR B0655+64, \citealt{Tauris00}). 
And, as noted above, the neutron star in J1909-3744
also appears to have spiraled through a massive common envelope.
However, it seems possible that systems with massive
white-dwarfs secondaries could also be produced via the double-core channel.
We note that, even though the double-core channel requires very
special initial conditions, such systems are more compact when the
first supernova occurs and are therefore much more likely to survive
as bound systems after the first supernova than in the standard
channel.  As a consequence, the birthrate of double neutron star systems
in the double-core channel can be
comparable to the standard model: using binary population synthesis
simulations, \citet{Dewi06} estimated their birthrate to be
$10^{-6}-10^{-5}\,$yr$^{-1}$ (but with substantial uncertainties, see  \citet{Belczynski06} for another estimate).
This is probably sufficient to account for the observed number of
double neutron-star systems (see, e.g., \citealt{Kalogera04}).
Nonetheless, whether or not neutron stars can
survive a common envelope, we believe we have provided some further
support for the double-core channel formation of double neutron star
systems.

\acknowledgments
We thank Chris Fryer, Michael Kramer, Onno Pols, and Ed van den Heuvel 
for extremely helpful discussions.


\end{document}

%% file: ns_masses.tbl
J0737-3039A/B	& $1.3381 \pm 0.0007$	& $1.2489 \pm 0.0007$	& 2.4	& 0.088			& 23			  & \citet{Kramer06}	  \\
B1534+12 	& $1.3332 \pm 0.0010$ 	& $1.3452 \pm 0.0010$  	& 10.1 	& 0.273 		& 38 			  & \citet{Stairs02}	  \\
J1756-2251 	& $1.32 \pm 0.02$ 	& $1.24 \pm 0.02$ 	& 7.67 	& 0.18 			& 28			  & \citet{Stairs08}	  \\
J1906+0746 	& $1.365 \pm 0.018$ 	& $1.248 \pm 0.018$ 	& 3.98 	& 0.085 		& 144\tablenotemark{\dag} & \cite{Kasian08}	  \\
B1913+16 	& $1.4414 \pm 0.0002$ 	& $1.3867 \pm 0.0002$ 	& 7.92 	& 0.617 		& 59			  & \citet{Weisberg05}	  \\
B2127+11C 	& $1.358 \pm 0.010$ 	& $1.354 \pm 0.010$ 	& 8.05 	& 0.681 		& 30			  & \citet{Jacoby06}	  \\
J1909-3744 	& $1.438 \pm 0.024$  &    white dwarf 	& 36.7 	& $\lesssim 10^{-6}$ 	& 2.9 			  & \cite{Jacoby05}	  \\
J1141-6545 	&    white dwarf  & $1.27  \pm 0.01$		& 4.74 	& 0.172     		& 393\tablenotemark{\dag} & \cite{Bhat08}

%% file: ns_masses.bbl
\begin{thebibliography}{45}
\expandafter\ifx\csname natexlab\endcsname\relax\def\natexlab#1{#1}\fi

\bibitem[{{Belczynski} {et~al.}(2009){Belczynski}, {Lorimer}, {Ridley}, \&
  {Curran}}]{Belczynski09}
{Belczynski}, K., {Lorimer}, D.~R., {Ridley}, J.~P., \& {Curran}, S.~J. 2009,
  ArXiv e-prints

\bibitem[{{Belczynski} {et~al.}(2006){Belczynski}, {Perna}, {Bulik},
  {Kalogera}, {Ivanova}, \& {Lamb}}]{Belczynski06}
{Belczynski}, K., {Perna}, R., {Bulik}, T., {Kalogera}, V., {Ivanova}, N., \&
  {Lamb}, D.~Q. 2006, \apj, 648, 1110

\bibitem[{{Bethe} \& {Brown}(1998)}]{Bethe98}
{Bethe}, H.~A., \& {Brown}, G.~E. 1998, \apj, 506, 780

\bibitem[{{Bhat} {et~al.}(2008){Bhat}, {Bailes}, \& {Verbiest}}]{Bhat08}
{Bhat}, N.~D.~R., {Bailes}, M., \& {Verbiest}, J.~P.~W. 2008, \prd, 77, 124017

\bibitem[{{Blondin} \& {Mezzacappa}(2006)}]{Blondin06}
{Blondin}, J.~M., \& {Mezzacappa}, A. 2006, \apj, 642, 401

\bibitem[{{Blondin} \& {Mezzacappa}(2007)}]{Blondin07}
---. 2007, \nat, 445, 58

\bibitem[{{Brandt} \& {Podsiadlowski}(1995)}]{Brandt95}
{Brandt}, N., \& {Podsiadlowski}, P. 1995, \mnras, 274, 461

\bibitem[{{Brown}(1995)}]{Brown95}
{Brown}, G.~E. 1995, \apj, 440, 270

\bibitem[{{Chevalier}(1993)}]{Chevalier93}
{Chevalier}, R.~A. 1993, \apjl, 411, L33

\bibitem[{{Chevalier}(1996)}]{Chevalier96}
---. 1996, \apj, 459, 322

\bibitem[{{Church} {et~al.}(2006){Church}, {Bush}, {Tout}, \&
  {Davies}}]{Church06}
{Church}, R.~P., {Bush}, S.~J., {Tout}, C.~A., \& {Davies}, M.~B. 2006, \mnras,
  372, 715

\bibitem[{{Cook} {et~al.}(1994){Cook}, {Shapiro}, \& {Teukolsky}}]{Cook94}
{Cook}, G.~B., {Shapiro}, S.~L., \& {Teukolsky}, S.~A. 1994, \apj, 422, 227

\bibitem[{{Dessart} {et~al.}(2006){Dessart}, {Burrows}, {Ott}, {Livne}, {Yoon},
  \& {Langer}}]{Dessart06}
{Dessart}, L., {Burrows}, A., {Ott}, C.~D., {Livne}, E., {Yoon}, S.-C., \&
  {Langer}, N. 2006, \apj, 644, 1063

\bibitem[{{Dewi} {et~al.}(2005){Dewi}, {Podsiadlowski}, \& {Pols}}]{Dewi05}
{Dewi}, J.~D.~M., {Podsiadlowski}, P., \& {Pols}, O.~R. 2005, \mnras, 363, L71

\bibitem[{{Dewi} {et~al.}(2006){Dewi}, {Podsiadlowski}, \& {Sena}}]{Dewi06}
{Dewi}, J.~D.~M., {Podsiadlowski}, P., \& {Sena}, A. 2006, \mnras, 368, 1742

\bibitem[{{Foglizzo} {et~al.}(2007){Foglizzo}, {Galletti}, {Scheck}, \&
  {Janka}}]{Foglizzo07}
{Foglizzo}, T., {Galletti}, P., {Scheck}, L., \& {Janka}, H.-T. 2007, \apj,
  654, 1006

\bibitem[{{Fryer} \& {Young}(2007)}]{Fryer07}
{Fryer}, C.~L., \& {Young}, P.~A. 2007, \apj, 659, 1438

\bibitem[{{Hobbs} {et~al.}(2005){Hobbs}, {Lorimer}, {Lyne}, \&
  {Kramer}}]{Hobbs05}
{Hobbs}, G., {Lorimer}, D.~R., {Lyne}, A.~G., \& {Kramer}, M. 2005, \mnras,
  360, 974

\bibitem[{{Jacoby} {et~al.}(2006){Jacoby}, {Cameron}, {Jenet}, {Anderson},
  {Murty}, \& {Kulkarni}}]{Jacoby06}
{Jacoby}, B.~A., {Cameron}, P.~B., {Jenet}, F.~A., {Anderson}, S.~B., {Murty},
  R.~N., \& {Kulkarni}, S.~R. 2006, \apjl, 644, L113

\bibitem[{{Jacoby} {et~al.}(2005){Jacoby}, {Hotan}, {Bailes}, {Ord}, \&
  {Kulkarni}}]{Jacoby05}
{Jacoby}, B.~A., {Hotan}, A., {Bailes}, M., {Ord}, S., \& {Kulkarni}, S.~R.
  2005, \apjl, 629, L113

\bibitem[{{Janka} {et~al.}(2008){Janka}, {Marek}, {M{\"u}ller}, \&
  {Scheck}}]{Janka08}
{Janka}, H.-T., {Marek}, A., {M{\"u}ller}, B., \& {Scheck}, L. 2008, in
  American Institute of Physics Conference Series, Vol. 983, 40 Years of
  Pulsars: Millisecond Pulsars, Magnetars and More, ed. C.~{Bassa}, Z.~{Wang},
  A.~{Cumming}, \& V.~M. {Kaspi}, 369--378

\bibitem[{{Kalogera} {et~al.}(2004){Kalogera}, {Kim}, {Lorimer}, {Burgay},
  {D'Amico}, {Possenti}, {Manchester}, {Lyne}, {Joshi}, {McLaughlin}, {Kramer},
  {Sarkissian}, \& {Camilo}}]{Kalogera04}
{Kalogera}, V., {et~al.} 2004, \apjl, 601, L179

\bibitem[{{Kasian}(2008)}]{Kasian08}
{Kasian}, L. 2008, in American Institute of Physics Conference Series, Vol.
  983, 40 Years of Pulsars: Millisecond Pulsars, Magnetars and More, ed.
  C.~{Bassa}, Z.~{Wang}, A.~{Cumming}, \& V.~M. {Kaspi}, 485--487

\bibitem[{{Kitaura} {et~al.}(2006){Kitaura}, {Janka}, \&
  {Hillebrandt}}]{Kitaura06}
{Kitaura}, F.~S., {Janka}, H.-T., \& {Hillebrandt}, W. 2006, \aap, 450, 345

\bibitem[{{Kramer} {et~al.}(2006){Kramer}, {Stairs}, {Manchester},
  {McLaughlin}, {Lyne}, {Ferdman}, {Burgay}, {Lorimer}, {Possenti}, {D'Amico},
  {Sarkissian}, {Hobbs}, {Reynolds}, {Freire}, \& {Camilo}}]{Kramer06}
{Kramer}, M., {et~al.} 2006, Science, 314, 97

\bibitem[{{Lattimer} \& {Prakash}(2001)}]{Lattimer01}
{Lattimer}, J.~M., \& {Prakash}, M. 2001, \apj, 550, 426

\bibitem[{{Mezzacappa} {et~al.}(2007){Mezzacappa}, {Bruenn}, {Blondin}, {Hix},
  \& {Bronson Messer}}]{Mezzacappa07}
{Mezzacappa}, A., {Bruenn}, S.~W., {Blondin}, J.~M., {Hix}, W.~R., \& {Bronson
  Messer}, O.~E. 2007, in American Institute of Physics Conference Series, Vol.
  924, The Multicolored Landscape of Compact Objects and Their Explosive
  Origins, ed. T.~{di Salvo}, G.~L. {Israel}, L.~{Piersant}, L.~{Burderi},
  G.~{Matt}, A.~{Tornambe}, \& M.~T. {Menna}, 234--242

\bibitem[{{M{\"u}ther} {et~al.}(1987){M{\"u}ther}, {Prakash}, \&
  {Ainsworth}}]{Muther87}
{M{\"u}ther}, H., {Prakash}, M., \& {Ainsworth}, T.~L. 1987, Physics Letters B,
  199, 469

\bibitem[{{Nomoto}(1984)}]{Nomoto84}
{Nomoto}, K. 1984, \apj, 277, 791

\bibitem[{{Nomoto}(1987)}]{Nomoto87}
---. 1987, \apj, 322, 206

\bibitem[{{Nomoto} \& {Iben}(1985)}]{Nomoto85}
{Nomoto}, K., \& {Iben}, Jr., I. 1985, \apj, 297, 531

\bibitem[{{Nomoto} \& {Kondo}(1991)}]{Nomoto91}
{Nomoto}, K., \& {Kondo}, Y. 1991, \apjl, 367, L19

\bibitem[{{Pfahl} {et~al.}(2002){Pfahl}, {Rappaport}, {Podsiadlowski}, \&
  {Spruit}}]{Pfahl02}
{Pfahl}, E., {Rappaport}, S., {Podsiadlowski}, P., \& {Spruit}, H. 2002, \apj,
  574, 364

\bibitem[{{Podsiadlowski} {et~al.}(2005){Podsiadlowski}, {Dewi}, {Lesaffre},
  {Miller}, {Newton}, \& {Stone}}]{Podsi05}
{Podsiadlowski}, P., {Dewi}, J.~D.~M., {Lesaffre}, P., {Miller}, J.~C.,
  {Newton}, W.~G., \& {Stone}, J.~R. 2005, \mnras, 361, 1243

\bibitem[{{Podsiadlowski} {et~al.}(2004){Podsiadlowski}, {Langer},
  {Poelarends}, {Rappaport}, {Heger}, \& {Pfahl}}]{Podsi04}
{Podsiadlowski}, P., {Langer}, N., {Poelarends}, A.~J.~T., {Rappaport}, S.,
  {Heger}, A., \& {Pfahl}, E. 2004, \apj, 612, 1044

\bibitem[{{Poelarends} {et~al.}(2008){Poelarends}, {Herwig}, {Langer}, \&
  {Heger}}]{Poelarends08}
{Poelarends}, A.~J.~T., {Herwig}, F., {Langer}, N., \& {Heger}, A. 2008, \apj,
  675, 614

\bibitem[{Press {et~al.}(2007)Press, Teukolsky, Vetterling, \& Flannery}]{NR07}
Press, W., Teukolsky, S., Vetterling, W., \& Flannery, B. 2007, {Numerical
  Recipes: The Art of Scientific Computing} (Cambridge University Press)

\bibitem[{{Siess}(2007)}]{Siess07}
{Siess}, L. 2007, \aap, 476, 893

\bibitem[{{Stairs}(2008)}]{Stairs08}
{Stairs}, I.~H. 2008, in American Institute of Physics Conference Series, Vol.
  983, 40 Years of Pulsars: Millisecond Pulsars, Magnetars and More, ed.
  C.~{Bassa}, Z.~{Wang}, A.~{Cumming}, \& V.~M. {Kaspi}, 424--432

\bibitem[{{Stairs} {et~al.}(2002){Stairs}, {Thorsett}, {Taylor}, \&
  {Wolszczan}}]{Stairs02}
{Stairs}, I.~H., {Thorsett}, S.~E., {Taylor}, J.~H., \& {Wolszczan}, A. 2002,
  \apj, 581, 501

\bibitem[{{Tauris} {et~al.}(2000){Tauris}, {van den Heuvel}, \&
  {Savonije}}]{Tauris00}
{Tauris}, T.~M., {van den Heuvel}, E.~P.~J., \& {Savonije}, G.~J. 2000, \apjl,
  530, L93

\bibitem[{{van den Heuvel}(2004)}]{vdH04}
{van den Heuvel}, E.~P.~J. 2004, in ESA Special Publication, Vol. 552, 5th
  INTEGRAL Workshop on the INTEGRAL Universe, ed. {V.~Schoenfelder, G.~Lichti,
  \& C.~Winkler}, 185

\bibitem[{{van den Heuvel}(2006)}]{vdH06}
{van den Heuvel}, E.~P.~J. 2006, Advances in Space Research, 38, 2667

\bibitem[{{van den Heuvel}(2007)}]{vdH07}
{van den Heuvel}, E.~P.~J. 2007, in American Institute of Physics Conference
  Series, Vol. 924, The Multicolored Landscape of Compact Objects and Their
  Explosive Origins, ed. {T.~di Salvo, G.~L.~Israel, L.~Piersant, L.~Burderi,
  G.~Matt, A.~Tornambe, \& M.~T.~Menna}, 598--606

\bibitem[{{Weisberg} \& {Taylor}(2005)}]{Weisberg05}
{Weisberg}, J.~M., \& {Taylor}, J.~H. 2005, in Astronomical Society of the
  Pacific Conference Series, Vol. 328, Binary Radio Pulsars, ed. F.~A. {Rasio}
  \& I.~H. {Stairs}, 25

\end{thebibliography}
